\documentclass[prb,twocolumn,amsmath,amssymb,superscriptaddress]{revtex4}
\usepackage{graphicx}
\usepackage{dcolumn}
\usepackage{bm}

\newcount\bozza \bozza=1
\ifnum\bozza=1
\newdimen\shift \shift=-2truecm
\def\lb#1{%
{\label{#1}\rlap{\kern\shift{$\scriptstyle#1$}}}}
\else\def\lb#1{\label{#1}} \fi

\usepackage{amsmath}
\usepackage{amssymb}

\begin{document}



\title{Massless Dirac cones in graphene: experiments and theory}


\author{E. Cappelluti\footnote{Corresponding author.
Email: emmanuele.cappelluti@roma1.infn.it,
Ph.: +39-06-4993-7453, Fax: +39-06-4993-7440.
}}
\affiliation{Istituto dei Sistemi Complessi, U.O.S. Sapienza, CNR, Rome, Italy}
\affiliation{Dipartimento di Fisica, Universit\`a ``La Sapienza'', Rome, Italy}

\author{L. Benfatto}
\affiliation{Istituto dei Sistemi Complessi, U.O.S. Sapienza, CNR, Rome, Italy}
\affiliation{Dipartimento di Fisica, Universit\`a ``La Sapienza'', Rome, Italy}

\author{M. Papagno}
\affiliation{Istituto di Struttura della Materia, CNR, Trieste, Italy}
\affiliation{Dipartimento di Fisica, Universit\`a della Calabria, 
Arcavacata di Rende (CS), Italy}

\author{D. Pacil\`e}
\affiliation{Istituto di Struttura della Materia, CNR, Trieste, Italy}
\affiliation{Dipartimento di Fisica, Universit\`a della Calabria,
Arcavacata di Rende (CS), Italy}

\author{P.M. Sheverdyaeva}
\affiliation{Istituto di Struttura della Materia, CNR, Trieste, Italy}

\author{P. Moras}
\affiliation{Istituto di Struttura della Materia, CNR, Trieste, Italy}

\begin{abstract}
The opening of a gap in single-layer graphene is often
ascribed to the breaking of the equivalence between the two carbon sublattices.
We show by angle-resolved photoemission spectroscopy that Ir- and
Na-modified graphene
grown on the Ir(111) surface presents a very large unconventional gap
that can be described in terms of a phenomenological “massless” Dirac model. 
We discuss the consequences and differences of this model
in  comparison of the standard massive gap model,
and we investigate the conditions under which such anomalous gap
can arise from a spontaneous symmetry breaking.
\end{abstract}

\date{\today}


\maketitle

\section{Introduction}

The isolation of single-layer and few-layer graphene
has triggered a huge burst of interest,
mainly motivated by the observation of unconventional electronic
properties, which stem from the Dirac-like low-energy
electronic structure of graphene characterized by a
gapless conical dispersion [\onlinecite{review}].
The huge electronic mobility of free-standing graphene
originates from its chiral properties, tightly linked with
the lack of electron backscattering phenomena near
the Fermi level [\onlinecite{review}]. 
A drawback of this characteristic band structure is the absence of an energy gap
between the Dirac cones, which would be highly desirable for the
exploitation
of graphene in device applications.
So far, however,
the effective employment of graphene-based materials
in low-energy electronics has been
hindered by the difficulty of opening a bandgap
without affecting the electronic mobility.
Understanding the fundamental mechanisms responsible
of gap opening in graphene is thus of the highest importance
in the perspective of engineering new efficient switch on/off devices.

From the theoretical point of view,
the simplest way to open a gap in the conical Dirac-like
dispersion of a two-dimensional honeycomb material
is to induce an inequivalence between the two
carbon sublattices A and B. This corresponds to include
a $\propto \hat{\sigma}_z$ term in the Dirac-like Hamiltonian:
\begin{eqnarray}
\hat{H}
&=&
E_{\rm D}\hat{I}
+
\sum_{\bf k}
\hbar v\left[
k_x\hat{\sigma}_x
+
k_y\hat{\sigma}_y
\right]
+\frac{\Delta}{2}\hat{\sigma}_z,
\label{hgap}
\end{eqnarray}
where $v$ is the Dirac velocity,
${\bf k}$ is the momentum relative to the K point,
and
where $\hat{\sigma}_i$ are $2 \times 2$
Pauli matrices defined in the subspace of the two carbon orbitals
for unit cell.
The term $E_{\rm D}$ represents an
energy off-shift of the Dirac point due to finite doping.
The total dispersion results thus:
$E_{k,\pm}=E_{\rm D}\pm\sqrt{(\hbar v k)^2+\Delta^2/4}$,
where $k=|{\bf k}|$.
Note that this model predicts for small ${\bf k}$ a parabolic behavior
$E_{k,\pm} \approx E_{\rm D}\pm\left(\Delta/2 +\hbar^2 k^2/2m_{\rm eff}\right)$ 
for both
conduction and valence bands with an effective mass
$m_{\rm eff}$ proportional to the band gap $\Delta$:
$m_{\rm eff}=\Delta^2/2v$. In order to highlight the strict relation
between the gap opening and the onset of an effective mass,
we define this scenario as ``massive'' (ms) gap model,
in contrast with a ``massless'' (ml) gap model that we will
discuss below.
It should be worth to note, in addition, that, in the massive gap
model, the band dispersion recovers a standard linear behavior
($E_{k,\pm} \approx E_{\rm D}\pm \hbar v k$)
for $\hbar v k \gg \Delta/2$, so that the extrapolation
of the upper band dispersion overlaps
the lower band and vice versa.

Angle-resolved photoemission spectroscopy (ARPES)
is one of the most direct experimental
techniques that allow to investigate the energy-momentum dependence
of the electronic states in solids.
Although ARPES can be applied also to suspended graphene,
as recently demonstrated in Ref. [\onlinecite{knox}],
it is more often applied to study the electronic structure of
supported graphene.
When graphene is in contact with a solid material finite doping
effects, as well as modifications (screening)
of the many-body interactions, are observed.
Gap opening is frequently reported,
with magnitudes that depend on the growth process
and on the substrate [\onlinecite{lanzara1,lanzara2,
gaparpes1,gaparpes2,gaparpes3,gaparpes4,
gaparpes5,gaparpes6,gaparpes7,gaparpes8,
gaparpes9,gaparpes10,gaparpes11,gaparpes12,acsnano}].
These observations are
commonly interpreted in term of the massive model
discussed above.
At a closer look, however, the actual
evidence of such physical behavior
in most of the cases is not assessed.
A first issue regards the
effective opening of a gap in some epitaxially grown graphene samples,
where a spectral anomaly at
the Dirac point was shown to be associated
with the signature of plasmaronic subbands rather than
with a simple gap opening [\onlinecite{eli1,eli2,eli3,eli4,eli5,eli6}].
On the other hand, ARPES data for other supported graphene layers do
not show the characteristic ``diamond-shaped'' dispersion
of the plasmaron at the Dirac point [\onlinecite{lanzara1,lanzara2,acsnano}].
Even in these cases, however, the actual experimental band dispersion
presents many inconsistencies with the
simple massive Dirac gap model,
pointing thus towards different mechanism of gap opening.

The ARPES data of
Ref. [\onlinecite{lanzara1,lanzara2}], for instance, reported a gap
$\Delta=0.26$ eV which was initially discussed in terms
of the massive gap model, and critically revised
in Ref. [\onlinecite{bc}].
Two main unconventional features
were there pointed out in regards to the ARPES data:
($i$) in spite of a large gap opening,
the conduction
and valence bands in Ref. [\onlinecite{lanzara1,lanzara2}]
retained a conical shape,
in contrast with the expected parabolic behavior;
($ii$) the linear extrapolation of the upper and lower bands
appeared to be {\em misaligned}, with an energy shift
corresponding to the gap.

In order to account for these unconventional features,
an alternative scenario  was
proposed in Ref. [\onlinecite{bc}] in terms of a phenomenological
``massless'' gap model with band dispersion 
$E_{k,\pm}=E_{\rm D}\pm \left[\Delta/2+\hbar v k\right]$.
The anomalous features ($i$)-($ii$)
could be naturally reproduced assuming
a self-energy of the form
$\hat{\Sigma}_{\bf k}=\Delta
[\hat{\sigma}_x\cos\theta +\hat{\sigma}_y\sin\theta]$, where
$\theta=\arctan(k_y/k_x)$.
This phenomenological model was also shown
to provide precise predictions (e.g. on the density of states) which
could be experimentally checked [\onlinecite{bc}].

The aim of the present paper is twofold.
On one hand we would like to provide
experimental evidence
of the validity of the massless gap model in CVD grown graphene.
On the other hand we discuss in details at the theoretical level
the possibility of the appearing of a massless gap
as result of a tendency towards a second-order phase transition.
We believe that a full understanding and controlling
of the unconventional properties of the massless gap
in graphene on substrates can open new perspectives
in the bandgap engineering in graphene-based materials.

\section{Massive vs. massless model}

Before addressing a quantitative analysis
of the experimental ARPES dispersion, and
a deeper discussion about the possible origin
of the anomalous bandgap features there observed,
we summarize briefly here the comparison between
the massive and massless gap models.

We first write the non-interacting Hamiltonian
in the form:
\begin{eqnarray}
\hat{H}^0_{\bf k}
&=&
E_{\rm D}
+
\hbar v k\left(
\begin{array}{cc}
0 & \mbox{e}^{-i\theta} \\
\mbox{e}^{i\theta} & 0
\end{array}
\right),
\label{h0}
\end{eqnarray}
where ${\bf k}$ are the momenta relative
to the K point.
The eigenvalues of Eq. (\ref{h0}),
$E_k^\pm=E_{\rm D}\pm \hbar v k$, describe the Dirac cone dispersion,
while the non-trivial dependence of (\ref{h0})
on the angle $\theta$ accounts for the chiral properties of the
eigenstates.

A massive gap is induced if
the A and B sublattices are electrostatically
inequivalent, for instance as an effect of the substrate.
Such inequivalence is formally taken into account by
the additional term
\begin{eqnarray}
\hat{H}^\Delta
&=&
E_{\rm D}
+
\frac{\Delta}{2}\hat{\sigma}_z
=
E_{\rm D}
+
\left(
\begin{array}{cc}
\Delta/2 & 0 \\
0 & -\Delta/2
\end{array}
\right),
 \label{gap}
\end{eqnarray}
as in Eq. (\ref{hgap}), resulting
in the well-known dispersion
\begin{eqnarray}
E_{k,\pm}^{\rm ms}
&=&
E_{\rm D}\pm\sqrt{(\hbar v k)^2+\Delta^2/4}.
\label{egap}
\end{eqnarray}
Two regimes can be identified in this
energy dispersion.
At high-energies ($k \gtrsim \Delta/\hbar v$)
the effect of the gap is negligible and the band dispersion
recovers the normal linear behavior $E_{k,\pm} \approx E_{\rm D}\pm \hbar v k$.
At low-energies  ($k \lesssim \Delta/\hbar v$)
the band acquires a parabolic shape $E_{k,\pm}
\approx E_{\rm D}\pm \left[\Delta/2+\hbar^2k^2/2m_{\rm eff}\right]$,
with $m_{\rm eff}=\Delta^2/2v$.
It is worth noting  that in this regime the chiral structure of the
band dispersion
is strongly affected by the opening of the gap [\onlinecite{mucha}].

The onset of a massive gap
in the low-energy electronic structure of graphene
has been discussed not only as due to the interaction with
the substrate, but also as a possible effect of a spontaneous
excitonic instability induced by the long-range Coulomb
interaction
[\onlinecite{appel,khveshchenko1,khveshchenko2,vafek,juricic,drut,liu,gamayun,fertig1,gonzalez3,fertig2,kotov,gonzalez1,gonzalez2,wang,popovici}].
In this latter context, the term (\ref{gap}) should be regarded
as an {\em order parameter}, rather than an intrinsic property of the system.
For real graphene systems, however, the dimensionless
Coulomb coupling constant results to be much smaller than the critical
value required for the excitonic instability.
On the other hand, in the absence of such excitonic broken symmetry,
in the normal state the long-range Coulomb interaction
has been shown to be accounted , at the leading order,
by a self-energy term [\onlinecite{voz,mish}]:
\begin{eqnarray}
\hat{\Sigma}_{{\bf k},\rm Cou}
&\propto&
\left(
\begin{array}{cc}
0 & k\ln({\cal K}/k)\mbox{e}^{-i\theta} \\
k\ln({\cal K}/k)\mbox{e}^{i\theta} & 0
\end{array}
\right).
 \label{coulomb}
\end{eqnarray}

Note here the much weaker dependence on $k$,
than the bare Dirac Hamiltonian. Indeed, the conventional
linear behavior in $k$ is here counterbalanced by
a logarithmic divergence $\sim \ln(k)$ induced
by the long-range Coulomb interaction.
Note finally that in this case
the chiral structure is preserved down to $k \rightarrow 0$.

The massless gap model investigated in this paper,
alternative to the massive model described by Eq. (\ref{gap}),
resembles features from both (\ref{gap}) and (\ref{coulomb}),
and it can be mathematically described by a self-energy term [\onlinecite{bc}]:
\begin{eqnarray}
\hat{\Sigma}_{\rm ml}
&\propto&
\frac{\Delta}{2}
\left(
\begin{array}{cc}
0 & \mbox{e}^{-i\theta} \\
\mbox{e}^{i\theta} & 0
\end{array}
\right),
 \label{less}
\end{eqnarray}
where the off-diagonal terms do {\em not} scale to zero
for $k \rightarrow 0$, but they still preserve
the full chiral properties.
As mentioned above,
the characteristic energy-momentum dispersion
results to be
\begin{eqnarray}
E_{k,\pm}^{\rm ml}
&=&
E_{\rm D}\pm \left[\Delta/2+\hbar v k\right].
\label{eless}
\end{eqnarray}
Note that this models accounts in a simple way
for the anomalous features described in the
introduction, namely:
$i$) the Dirac cones appear to be just split by the gap $\Delta$
without affecting their conical shape; $ii$) the perfectly linear
behavior of the upper (lower) band does not extrapolate onto
the linear behavior of the lower (upper) band.

As we are going to see, such anomalies can permit
to identify in a clear way one gap model rather than
the other one in the experimental data.

\section{Massless gapped Dirac model:
an ARPES evidence}

In this section we present a detailed analysis of
some recent ARPES measurements on single-layer
graphene on Ir(111) decorated by Ir clusters and Na alkali metals.
We show how these measurements
display,  in correspondence with the opening of a large energy gap [\onlinecite{acsnano}],
an anomalous behavior of the Dirac cones
compatible with the massless gap model.

\begin{figure}[b]
\includegraphics[scale=0.3,clip=]{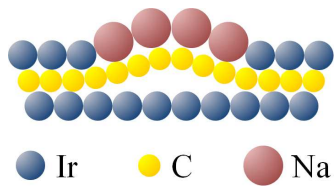}
\caption{Schematic sketch of the Na/Ir/G sample.}
\label{f-sketch}
\end{figure}

The experiment was performed at the VUV-Photoemission beamline
on the Elettra storage
ring in Trieste under ultrahigh vacuum conditions.
The Ir(111) crystal was cleaned by cycles of Ar$^+$ ion sputtering and
annealing at $T = 1500$ K. Surface order and cleanliness
of the sample were
checked by low-energy electron diffraction and core level
photoemission measurements.
Graphene was grown by thermal cracking of ethylene (C$_2$H$_4$)
on the Ir substrate held at 1300 K for an exposure of 100 L
(1L corresponds to an exposure of $10^{-6}$ mbar for 1s).
Under these experimental conditions, graphene had lattice vectors
aligned in plane
to those of the substrate and displayed 
a Moir\'e pattern originating from the interface lattice mismatch [\onlinecite{x1}].
Tiny Ir amounts were evaporated from a current heated thin film plate
($0.5$ mm width, $0.1$ mm thickness) at
an evaporation rate of about $2.0 \times 10^{-4}$ ML/s,
as determined by core level photoemission
measurements [\onlinecite{acsnano,x1}].
Ir evaporation was performed at a temperature of 350 K.
The deposition of Ir results in the nucleation of size selected Ir
clusters on the hcp regions of the Moir\'e supercells
(see Fig. \ref{f-sketch} for a schematic sketch) [\onlinecite{x1}]. 
The process was
finely controlled to saturate the hcp regions
of the Moir\'e-derived graphene supercells with one Ir cluster and to avoid
cluster percolation.
The optimal Ir coverage was determined to be 0.15 ML (monolayer) Ir,
with reference to the density
of an Ir(111) plane [\onlinecite{acsnano,x1}].
Additional Na was evaporated from a commercial getter source at
room temperature and
adsorbed to fill the residual
exposed graphene surface.
The resulting system (Na/Ir/G) was characterized by means of
angle-resolved photoemission,
as reported in Ref. [\onlinecite{acsnano}].
The valence band photoemission analysis was carried out
by a Scienta R4000 electron energy analyzer with $120$ eV photons,
close to the Cooper minimum of the Ir5d levels.
The spectra were collected at 120 K with energy and angular resolution
of 30 meV and 0.3$^\circ$, respectively.

\begin{figure}[t]
\includegraphics[scale=0.85,clip=]{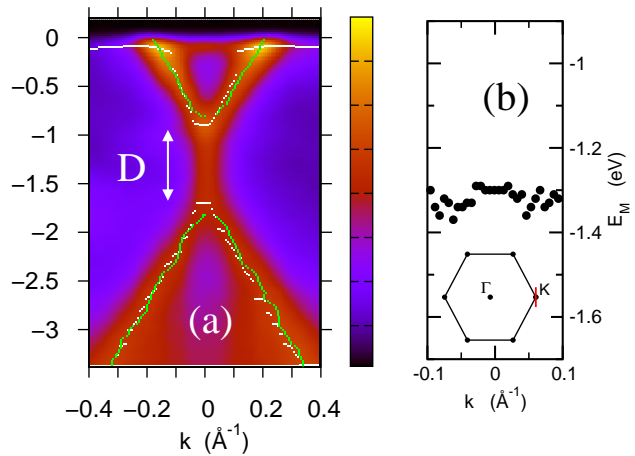}
\caption{(a) ARPES intensity map of the electronic
dispersion of Na/Ir/G along the direction p$\Gamma$K.
The light green points represent the electronic dispersion
determined by the momentum distribution curves, while
white points the electronic dispersion obtained by energy distribution
curves.
The vertical white double arrow marks the magnitude
$\Delta$ of the gap.
The location of the cut p$\Gamma$K within the Brillouin zone is shown
in the inset of panel (b).
(b) Momentum dependence of the average $E_{\rm M,k}$ (see text).}
\label{f-map}
\end{figure}

The ARPES intensity map, along the direction perpendicular to
$\Gamma$K passing through the K point (p$\Gamma$K)
is shown in Fig. \ref{f-map}a,
where the white and green symbols represent the position of the 
maxima
in the energy-distribution-curves (EDCs)
and in the momentum-distribution-curves (MDCs), respectively.
The Fermi momentum along the p$\Gamma$K direction
is $k_{\rm F} \approx 0.19$ \AA$^{-1}$ that,
taking into account the trigonal warping of the bands, gives
an electron doping of $n\approx 0.028$ electrons per C atom.
From the EDC at ${\bf k}=0$ we estimate a bandgap $\Delta \approx 0.8$ eV.

Two remarkable features stand out from
the ARPES data: ($i$) both conduction and valence bands
present a linear dispersion throughout the probed ${\bf k}$-space region,
except within a very small region $k \lesssim 0.026$ \AA$^{-1}$.
This range is thus much
smaller than that predicted by the massive
gap model ($k \lesssim \Delta/\hbar v \approx 0.075$ \AA$^{-1}$);
($ii$) the linear extrapolations of the upper and lower branched do
not match,
at variance with the expectations of the massive gap model.
We note additionally the lack of any 
diamond-like features close to the K point,
which would be typical trademarks of plasmaronic
coupling [\onlinecite{eli1,eli2,eli3,eli4,eli5,eli6}].
On the other hand,
plasmaron signatures in ARPES data have been so far
reported only for epitaxial graphene on SiC, whereas
they are absent in graphene grown
on metallic substrates, most probably as a consequence of
screening effects.

Such band structure of the Na/Ir/G system appears to be
incompatible with the standard massive gap model,
whereas they could be naturally reproduced within
the context of the massless gap model.

In order to address at a more quantitative level this issue,
we provide here a detailed careful analysis of the ARPES
data for Na/Ir/G.
Before testing the different models,
it is necessary to estimate the basic parameters
which are {\em independent} of the model itself,
e.g. the Fermi velocity $v$ and the the Dirac energy $E_{\rm D}$
in the absence of the gap.

We estimate $\hbar v$ from the 
linear behavior of the lower band in the energy
window from $-3.5$ eV to $-1.7$ eV, which yields
$\hbar v \approx 5.3$ eV/\AA.
This value is consistent with the speed value derived
from the upper band.
We found this value was also perfectly compatible
with the linear slope of the upper band.
A rough estimate $E_{\rm D}=-1.3$ eV
of the Dirac energy 
in the absence of the gap is obtained as the midgap point
at ${\bf k}=0$
($E_{k=0,+}=-0.9$ eV, $E_{k=0,-}=-1.7$ eV).
This estimate can be cross-checked with the average energy between
the upper and lower band
$E_{\rm M,k}=(E_{k=0,-}+E_{k=0,-})/2$
for generic ${\bf k}$, displayed in Fig. \ref{f-map}b.
The result
shows a negligible momentum dependence, resulting
in $E_{\rm M,k}=-1.3$ eV.
Note that the negligible dependence of $E_{\rm M,k}$ on ${\bf k}$
signalizes an almost perfect symmetry of the experimental
upper and lower bands
with respect to the Dirac energy. 
This is compatible with both the massless and massive gap models,
but it rules out plasmaronic effects which would result
in an asymmetry and in a finite ${\bf k}$ dependence of $E_{\rm M,k}$.

The unbiased determination of $E_{\rm D}$ and $v$
allows to test the validity of the different models on the
experimental data.
\begin{figure}[t]
\includegraphics[scale=0.5,clip=]{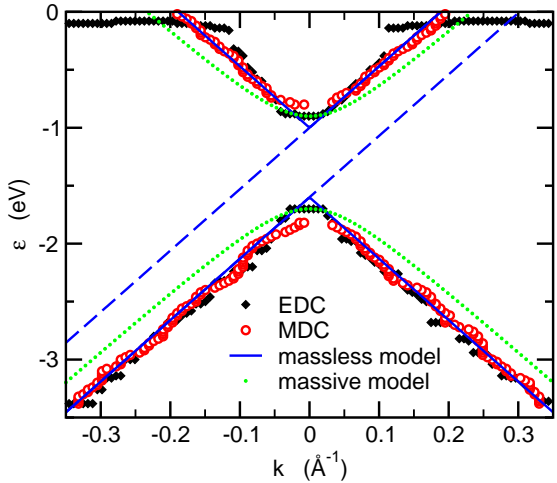}
\caption{Comparison between the experimental dispersion for Na/Ir/G
and the model predictions. Black filled diamonds and red empty circles
represent EDC and MDC data, respectively.
Solid blue lines the predictions of the massless model
with $\Delta=0.3$ eV, and blue dashed lines the
corresponding extrapolations of the Dirac cones.
The dotted green lines show the massive gap model
with $\Delta=0.4$ eV. In both theoretical cases we have used
$E_{\rm D}=-1.3$ eV and $\hbar v \approx 5.3$ eV/\AA.}
\label{f-disp}
\end{figure}
Fig. \ref{f-disp}
compares the band dispersion extracted from the momentum and energy
distribution curves with the prediction of massless and massive modes. The best agreement is found
for the massless model with $\Delta=0.3$ eV.
Note the mismatch between the upper and lower Dirac cones,
that reproduce perfectly the experimental data.
The same value of the gap can be indeed also inferred from such
misalignment.
On the other hand,
no satisfactory fit was possible for the massive model.
We show here thus the predictions
of the massive model for $\Delta=0.4$ eV, that reproduces the value
at K of the upper and lower bands.

\section{Microscopic models}

In the previous section we have shown how a careful analysis
of the band anomalies observed by ARPES in in Na/Ir/G 
points out towards a phenomenology of the graphene bands
compatible more with massless gap model
rather than with the standard massive one.
It is clear that the different models represent a trademark
for different underlying physical processes.
In this Section we address thus this issue, in order to clarify which
kind of physical processes can be responsible
at the microscopic level for the massless Dirac phenomenology,
with the final aim to control and engineering such processes.

Along this perspective, a brief summary of the microscopic physics
related to the massive model can be useful.
As mentioned in the Introduction, the massive Dirac model
essentially stems from the presence of a local potential
term $\propto \hat{\sigma}_z$ which makes the two carbon atom
sublattices inequivalent.
This potential can be naturally induced by the interaction of graphene with the substrate.
The term $\Delta \hat{\sigma}_z$
in the Hamiltonian appears then as an {\em external} field
arising from the environment.
Alternatively, the onset of a mass  term
$\propto  \hat{\sigma}_z$ has been discussed within the context
of a spontaneous symmetry breaking, where
the driving mechanism is the unscreened long-range Coulomb
interaction [\onlinecite{appel,khveshchenko1,khveshchenko2,vafek,juricic,drut,liu,gamayun,fertig1,gonzalez3,fertig2,kotov,gonzalez1,gonzalez2,wang,popovici}].
The term $\Delta({\bf k}) \hat{\sigma}_z$ can be viewed
thus in this scenario as an anomalous {\em self-energy} term,
i.e. as an {\em order parameter}, arising
from many-body effects.
It should be however mentioned that in this case
the mass term $\Delta({\bf k})$
appears to be momentum dependent,
i.e. divergent for  ${\bf k} \rightarrow 0$
[\onlinecite{gamayun,khveshchenko1}], rather than
${\bf k}$-independent.

From a general point of view,
since the presence of a linear Dirac cone
in ideal flat graphene is dictated by symmetry reasons,
it is clear that
the opening of a gap and the splitting of the upper and lower bands,
both in the massive and in the massless model,
must be associated with some kind of breaking symmetry.
We have indeed just seen how the symmetry breaking
(spontaneous or not) responsible for
the massive model is the equivalence between the
local potentials of the two sublattices.
In Ref. [\onlinecite{bc}] the general constraints that a
phenomenological self-energy must satisfied to
give rise to a massless Dirac model have been discussed.
It is on the other hand intriguing to discuss in the present work
the conditions that could make possible the appearance
of a self-energy (\ref{less}) as result of a tendency towards a spontaneous symmetry
breaking.

It is clear that
the effective breaking of symmetry, in real materials, can be
prevented by many causes. For instance, as we will discuss below,
a good candidate for a quasi-breaking of symmetry is
the long-range Coulomb interaction. In real materials,
and in particular in Na/Ir/graphene, such interaction is expected to
be screened resulting in a cut-off of the singular behavior
for small exchange momenta ${\bf q}$.
Asymptotically, this should correspond to ``regularization''
of the interaction, and thus to a linear behavior of the dispersion
(although with a very steep slope) for momenta ${\bf k}$ close to the K point.
For larger ${\bf k}$, however, the tendency towards an instability can still
be reflected in a quasi-gapped dispersion which can be compatible
with the experimental observations.
From this point of view, we present the analysis of the onset of a
spontaneous symmetry breaking just as an illustrative example
for the gross features of the resulting phenomenology.
In similar way, we do not want to provide here a quantitative evaluation
of the possible critical couplings required to induce a massless
gapped case, rather to compare, at a qualitative level, this strength
of this instability with the strength of the corresponding instability
associated with the spontaneous
generation of a mass according the model (\ref{hgap}).
In order to have thus a direct and clear way to compare the two
possible instabilities, we address here this issue
within the context of a mean-field solution, keeping in mind
that a more compelling high-order analysis is needed
for a quantitative investigation.

In order to address this issue in the most convenient way,
we introduce the Nambu notation where the
Hamiltonian of ideal free-standing graphene
can
be written as:
\begin{eqnarray}
H
&=&
\sum_{{\bf k},\sigma}
\psi_{{\bf k},\sigma}^\dagger
\hat{H}_0({\bf k})
\psi_{{\bf k},\sigma}
,
\label{H0}
\end{eqnarray}
where $\sigma$ is a global index taking into
account both spin and valley (degenerate) degrees of freedom $N_s=2$, $N_v=2$),
$\psi_{{\bf k},\sigma}^\dagger$ is the spinor
in the sublattice basis 
$\psi_{{\bf k},\sigma}^\dagger=(c_{{\bf k},\sigma,\rm A}^\dagger,c_{{\bf k},\sigma,\rm B}^\dagger )$,
and where
\begin{eqnarray}
\hat{H}_0({\bf k})
&=&
\hbar vk\left[
\hat{\sigma}_x
\cos\theta
+
\hat{\sigma}_y \sin\theta
\right].
\end{eqnarray}

We assume
a long-range Coulomb interaction
which can we written in the momentum space as:
\begin{eqnarray}
H_{\rm C}
&=&
\frac{1}{2}
\sum_{\sigma,\sigma'}
\sum_{\bf k,p,q}
\rho_{\bf -q}
V({\bf q})
\rho_{\bf q},
\label{Hc}
\end{eqnarray}
where
$\rho_{\bf q}=\sum_{{\bf k},\sigma}\psi_{{\bf k+q},\sigma}^\dagger
\psi_{{\bf k},\sigma}$,
and where
$V({\bf q})=2\pi e^2 [1-\delta({\bf q})]/\epsilon_0 \kappa  |{\bf q}|$.
The term $\propto [1-\delta({\bf q})]$ in $V({\bf q})$
takes into account the subtraction of the positive charged
background, while $\kappa$ is the dimensionless relative
in-plane dielectric constant.

The phenomenology of the normal state self-energy
arising from such many-body interaction,
as well as of several possible {\em anomalous}
self-energies associated with different spontaneous symmetry breaking,
has been discussed in an extended way using different accurate
techniques, as 
for instance renormalization group (RG) or static random-phase
approximation (RPA). In general, in order to have an accurate estimate
of the {\em magnitude} of the corresponding self-energies
(and of the minimum strength of the interaction in the case
of spontaneous symmetry breaking), a self-consistent
screening of the quasi-particle excitations and of the
the effective screened Coulomb interaction is required.

On the other hand, a
qualitative insight on the matrix/momentum structure of
the possible anomalous self-energies is already
possible employing a simple mean-field treatment.

Neglecting the Hartree term, which does not play any role
in the present context, and focusing on the exchange (Fock) term,
we can write the generic self-energy as:
\begin{eqnarray}
\hat{\Sigma}
({\bf k})
&=&
T
\sum_{{\bf p},n}
\frac{V({\bf k-p})}
{i\omega_n\hat{I}
-\hat{H}_0({\bf k})-
\hat{\Sigma}({\bf k})
}
.
\label{fock}
\end{eqnarray}
Note that, since the interaction $V({\bf q})$
is peaked at small ${\bf q=k-p}$, the
intra-valley scattering is dominant, supporting the validity
of the low-energy model versus a full tight-binding treatment.
Note also that at this level of approximation
(unscreened static Coulomb interaction)
the self-energy does not depend on the frequency $i\omega_n$.
Within this scheme is thus convenient to define
an effective mean-field hamiltonian
$\hat{H}({\bf k})=\hat{H}_0({\bf k})+\hat{\Sigma}({\bf k})$,
and to write a recursive equation:
\begin{eqnarray}
\hat{H}({\bf k})
&=&
\hat{H}_0({\bf k})
+
T
\sum_{{\bf p},n}
\frac{V({\bf k-p})}
{i\omega_n\hat{I}
-\hat{H}({\bf k})
}.
\label{heff}
\end{eqnarray}

Since $V({\bf k-p})$ depends on the angular variables
$\theta_k$, $\theta_p$ essentially through only $\cos(\theta_k-\theta_p)$,
one can see that $\hat{H}({\bf k}) \approx
\beta(k)
[\hat{\sigma}_x\cos\theta+\hat{\sigma}_y \sin\theta$.
Apart numerical prefactor,
the momentum dependence of $\beta(k)$ on $k$
can be caught already by a perturbative analysis.
Replacing $\hat{H}({\bf k})$ with $\hat{H}_0({\bf k})$
in the right side term of Eq. (\ref{heff}), or, equivalently,
neglecting $\hat{\Sigma}({\bf k})$
in the right side term of Eq. (\ref{fock}),
we obtain:
\begin{eqnarray}
\beta(k)
&=&
\hbar v k
\left[
1+
\frac{\alpha}{2\kappa} \gamma(k)
\right]
\label{alpha}
\end{eqnarray}
where $\alpha=e^2/\epsilon_0 \hbar v$ is the coupling constant
for suspended graphene, ${\cal K}$ is a high-momentum cut-off limiting the
validity of the Dirac model, approximately determined
by the size of the Brillouin zone, and where
\begin{eqnarray}
\gamma(k)
&=&
\int_0^{{\cal K}/k} dx
\int_0^{2\pi}\frac{d\theta}{2\pi}
\frac{x \cos\theta}
{\sqrt{1+x^2-2x\cos\theta}}.
\label{emish}
\end{eqnarray}

A careful analysis of (\ref{emish}) shows that
at the leading order $\gamma(k)\approx (1/2)\ln({\cal K}/k)+$const.,
a well known result.
Note that, within this context, the self-energy associated with
the massless gap model is formally described by 
$\beta(k)=$const., i.e. by $\gamma(k)=1/k$.
It is clear that such analytical dependence is not spontaneously
generated by the self-consistent solution of (\ref{heff})
in the normal state, and it must be regarded as an order parameter
of a possible second order phase transition.
In order to address this possibility,
we rewrite Eq. (\ref{heff}) as:
\begin{eqnarray}
\hat{H}({\bf k})
-\hat{W}[{\bf k},\{\hat{H}({\bf p})\}]
&=&
\hat{H}_0({\bf k})
,
\label{hself}
\end{eqnarray}
where $\hat{W}[{\bf k},\{\hat{X}\}]$ is the functional
$\hat{W}[{\bf k},\{\hat{X}\}]=T
\sum_{{\bf p},n}
V({\bf k-p})/
[i\omega_n\hat{I}-\hat{X}]$.
Broken symmetry phases can be generated
when the equation
\begin{eqnarray}
\hat{\phi}({\bf k})
-\hat{W}[{\bf k},\{\hat{H}({\bf p})+\hat{\phi}({\bf p})\}]
&=&
0
\label{inst}
\end{eqnarray}
admits a non-trivial solution $\hat{\phi}({\bf k})\neq 0$.
We are here interested only in investigating the instability
of the normal state towards a broken symmetry phase.
Within this context we can expand thus
the right side term of Eq. (\ref{inst})
at the {\em linear} order
with respect to $\phi({\bf k})$.
At $T=0$
we get thus the
susceptibility equations:
\begin{eqnarray}
\hat{\phi}({\bf k})
&=&
\sum_{\bf p}
V({\bf k-p})
\frac{\hat{\phi}({\bf p})}
{2E({\bf p})},
\label{lastgap}
\end{eqnarray}
where $E({\bf p})={\rm det}[\hat{H}({\bf p})]$ is the energy dispersion
in the normal state.
Within the spirit of a mean-field approach we can neglect
in $E({\bf p})$ the self-energy
many-body effects driven by the Coulomb interactions,
and approximate
$E({\bf p}) \simeq {\rm det}[\hat{H}_0({\bf p})]=\hbar v |{\bf p}|$.
Eq. (\ref{lastgap})
can be used to investigate the instability towards both
a massive as well as massless gap.
In the first case $\hat{\phi}({\bf k}) \propto \hat{\sigma}_z$,
which clearly breaks the symmetry of the original Hamiltonian,
while in the second case
$\hat{\phi}({\bf k}) \propto \hat{\sigma}_x, \hat{\sigma}_y$.
Since terms $\propto \hat{\sigma}_x, \hat{\sigma}_y$
are already present in the normal state, it is not
straightforwardly apparent
at this level the nature of the symmetry breaking.
On this regards
it should be noticed that
in the massless gap model the symmetry breaking
is more specifically associated
with the anomalous ${\bf k}$-dependence
of the order parameter, in similarity with ${\bf q}=0$
flux phases discussed in different contexts, rather than
with the Pauli structure.

To make clearer this point, we employ the massless gap model
$\phi_{\rm ml}({\bf k})=\Delta_{\rm ml}
[\hat{\sigma}_x\cos\theta+\hat{\sigma}_y\sin\theta]$
as an ansatz for the {\em right} side of Eq. (\ref{lastgap}).
It is easy to check that the equations for
$\hat{\sigma}_x$, $\hat{\sigma}_y$ result to be
decoupled and degenerate.
From a careful analysis of the angular variable,
it is also straightforward to check
that the resulting $\phi({\bf k})$ in the left side of
Eq. (\ref{lastgap}) appears to be indeed
$\phi({\bf k})=\beta_{\rm ml}(k)
[\hat{\sigma}_x\cos\theta+\hat{\sigma}_y\sin\theta]$,
where the prefactor $\beta_{\rm ml}(k)$ will be 
determined by the effective self-consistency of
Eq. (\ref{lastgap}).
Using the explicit expression for the Coulomb interaction,
we obtain thus:
\begin{eqnarray}
\beta_{\rm ml}(k)
&=&
\Delta_{\rm ml} 
\frac{\alpha}{2\kappa}
\nonumber\\
&&\times
\int_0^{\cal K} dp
\int_0^{2\pi}d\theta
\frac{\cos\theta}
{\sqrt{k^2+p^2-2kp\cos\theta}}.
\label{ml2}
\end{eqnarray}

It is remarkably relevant to notice that,
contrary to Eq. (\ref{emish}),
the integral over $p$ is well behaved
for both $k \rightarrow 0$ and ${\cal K}\rightarrow \infty$.
In this limits we can thus properly look for the
self-consistent solution $\beta_{\rm ml}(k)=\Delta_{\rm ml}$.
We obtain thus the susceptibility equation:
\begin{eqnarray}
1
&=&
\frac{\alpha}{2\kappa}
\chi_{\rm ml}
\label{chi}
\end{eqnarray}
where 
\begin{eqnarray}
\chi_{\rm ml}
&=&
\int_0^\infty dx
\int_0^{2\pi}
\frac{d\theta}{2\pi}
\frac{\cos\theta}
{\sqrt{1+x^2-2x\cos\theta}}
\nonumber\\
&=&
\left[2-\sqrt{3}/2 \right] \simeq 1.13.
\label{chi2}
\end{eqnarray}
For free-standing graphene with $\kappa=1$
we can predict thus a finite critical coupling $\alpha_c^{\rm ms} \simeq 1.76$
above which the system undergoes spontaneously a transition
towards a massless gap phase.

We should stress that such value of $\alpha_c^{\rm ms}$
has been derived within the lowest order approximation,
and a more compelling analysis is needed
for a reliable quantitative estimate.
It is however instructive to compare such value
with the critical coupling required to induce,
{\em at the same level of approximation},
a spontaneous generation of mass.
Such instability can be determined by using the ansatz
$\phi_{\rm ms}({\bf k})=[\Delta_{\rm ms}/\sqrt{k}]\hat{\sigma}_z$
in Eq. (\ref{lastgap}).
After few straightforward steps,
we obtain the susceptibility equation
\begin{eqnarray}
1
&=&
\frac{\alpha}{2 \kappa}
\chi_{\rm ms}
\label{chims}
\end{eqnarray}
where 
\begin{eqnarray}
\chi_{\rm ms}
&=&
\int_0^\infty \frac{dx}{\sqrt{x}}
\int_0^{2\pi}
\frac{d\theta}{2\pi}
{\sqrt{1+x^2-2x\cos\theta}}
\nonumber\\
&=&\frac{\Gamma^4(1/4)}{4\pi^2} \simeq 4.38,
\label{chi2ms}
\end{eqnarray}
corresponding, for $\kappa=1$, to the well-known result
$\alpha_c^{\rm ms} \simeq 0.46$ [\onlinecite{gamayun,popovici}].

Although the critical coupling $\alpha_c^{\rm ml}$ for the spontaneous generation of
massless gap results to be larger than the critical coupling
for the massive gap, it remains still of the same order of magnitude
of $\alpha_c^{\rm ms}$ within the same level of approximation.
We know however that $\alpha_c^{\rm ms}$ is significantly
increased when higher order renormalization
effects (dynamical screening, self-energy effects, \ldots) are taken
into account, with possible $\alpha_c^{\rm ms}$ as large as
$\alpha_c^{\rm ms}\approx 2$, of the same order thus
of the present estimate of  $\alpha_c^{\rm ml}$.
Becomes thus crucial, in order to assess the dominant instability of
the normal state, to include high-order effect also for the massless
gap case, to permit a quantitative comparison.
In addition, it should not be excluded a possible interplay,
in the broken symmetry phase, between the two order parameters,
where one could favor and make more stable the other one.
This scenario is suggested by the simple analysis of the experimental data
of Fig. \ref{f-disp}, where the massless gap model reproduce well
the experimental dispersion in a broad energy range, but where
a finite parabolic behavior is nevertheless present at very small $k$.
We have already discuss how such weak parabolic behavior cannot
account in any case for the large observed gap.
The experimental dispersion could be thus compatible
with the possible coexistence of the both qualitatively different
gaps, with the massless (larger) one determining the high-energy range,
and the massive (smaller) one accounting for the residual parabolic
behavior at small $k$.
Further analysis in this direction is however needed to
investigate this issue.

In conclusion, in this work we have provided,
using angle-resolved photoemission spectroscopy,
a compelling evidence of the presence of an unconventional
gap in CVD grown graphene, properly described by
a massless gap model.
We have further investigated, on a theoretical ground,
the microscopic conditions that can give rise to a spontaneous
phase transition towards such massless gapped phase.
We have shown that, for suspended graphene, a critical
value of the Coulomb coupling is needed,
with larger but comparable values with the ones required
to induce a breaking symmetry towards a massive map.
This analysis, stricty valid for suspended graphene,
can suggest a possible path to induce such unconventional gap
also in CVD graphene, although in this case specific features
of these materials
(screening of the Coulomb interaction, interference with the Moir\'e pattern)
must be taken into account.
The fully understanding and controlling
of the unconventional properties of the massless gap
in graphene can open new perspectives
in the bandgap engineering in graphene-based materials.

{\bf Acknowledgements}

E.C. acknowledges support from the European
project FP7-PEOPLE-2013-CIG "LSIE\_2D" and Italian
National MIUR Prin project 20105ZZTSE.
L.B. acknowledges support from
the Italian MIUR under the projects
FIRB-HybridNanoDev-RBFR1236VV and
PRIN-2012X3YFZ2.
Financial support was also provided by 
Italian MIUR project FIRB-Futuro in Ricerca 2010-Project
Plasmograph Grant No. RBFR10M5BT.

\end{document}